\documentclass[journal]{IEEEtran}

\usepackage{amsmath,amssymb,amsfonts}
\usepackage{graphicx}
\usepackage{xcolor}
\usepackage{multicol}
\usepackage{cite}

\newcommand{\T}{\ensuremath{\mathsf{T}}}

\newcommand{\bF}{\mathbf{F}}
\newcommand{\bG}{\mathbf{G}}
\newcommand{\bbi}{\mathbf{b}}
\newcommand{\bc}{\mathbf{c}}

\begin{document}

\title{On the Realization of Impulse Invariant Bilinear \\ Volterra Kernels}

\author{Phillip M.~S.~Burt and José Henrique de Morais Goulart
\thanks{P.~M.~S.~Burt is with \textit{Escola Politécnica, Universidade de São Paulo}, São Paulo, Brazil. e-mail: pmsburt@.usp.br. J.~H.~de M.~Goulart
is with \textit{IRIT, Université de Toulouse, Toulouse INP, CNRS, }
Toulouse, France. e-mail: henrique.goulart@irit.fr}}

\maketitle

\begin{abstract}
As previously shown, the direct extension of the impulse invariance principle to Volterra kernels has to be modified in order to provide a condition for the exact modeling of mixed-signal chains. At first sight this would seem to seriously complicate the otherwise simple discrete-time realization of separable kernels (among which bilinear kernels are of particular importance).  We show here, however, that this not the case. By defining a cascade operator, the structure of  a generalized impulse invariance can be unveiled, leading to a realization without an inordinate increase in computational complexity. \end{abstract}


\section{Introduction}

\IEEEPARstart{T}{he} impulse invariance \cite{Oppenheim1989} between discrete-time and continuous-time  linear time-invariant (LTI) systems provides a condition for the exact modeling of mixed-signal chains consisting of  discrete-time and continuous-time parts, such as in acoustic echo cancellation (AEC) \cite{Azpicueta-Ruiz2011}, for instance.

If the signal chain requires a nonlinear model, one might expect that a direct extension of the impulse invariance principle to Volterra kernels \cite{Rugh1981} would provide the condition for the same kind of modeling. In the case of separable kernels (among which bilinear kernels are of particular importance), this would be very convenient for their discrete-time realization, which would follow directly from their continuous-time realization.

It turns out, however, that when using triangular Volterra kernels (as required to minimize the resulting computational cost), the direct extension of the impulse invariance definition has to be somewhat modified in order to provide a condition for exact modeling, as pointed out in \cite{Rugh1981, Burt2018}. 
In this paper, we show that, as a consequence, even if the continuous-time kernels are separable, the resulting discrete-time ones are not.
At first sight this would seem to pose a serious problem to their realization. 
We show, though, that this is not the case, by describing how a separable kernel realization can be modified in order to implement the modified impulse invariance principle. To the best of our knowledge this problem has not been previously addressed.
Finally, we quantify the additional cost brought by this modification in number of operations.

This work is organized as follows. In Section \ref{sec:LIT} we revise the concept of LTI impulse invariance, its relation to modeling and realization aspects.  In Section \ref{sec:gen} we revise the generalization of impulse invariance to triangular Volterra kernels and reformulate it in terms of regular Volterra kernels. In Section \ref{sec:realiz} we show how impulse invariant separable kernels can be realized. Finally, in Section \ref{sec:complex} we assess the computational complexity of such realization.

\section{Impulse invariance of LTI systems}

\label{sec:LIT}

Given an LTI system with impulse response $h_c(t)$ and a sampling period $T$, the associated impulse-invariant \cite{Oppenheim1989} discrete-time system has impulse response \begin{equation} h(n)=h_c(nT). \label{eq:invariance} \end{equation} 

\subsection{Invariance and modeling}

\begin{figure}
    \centering
    \includegraphics{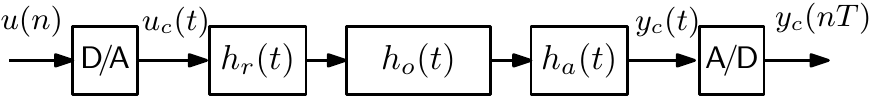}
    \caption{Signal chain starting in discrete-time, passing through continuous-time and then returning to discrete-time.}
    \label{fig:chain}
\end{figure}

The relation \eqref{eq:invariance} appears, for instance, when modeling a signal chain as depicted in Fig.~\ref{fig:chain}. The signal chain contains an ideal impulsive D/A providing $u_c(t)=\sum_n u(n)\delta(t-nT)$, a reconstruction filter $h_r(t)$, an LTI system $h_o(t)$, an anti-aliasing filter $h_a(t)$ and an A/D sampler. Its output then reads
\begin{align}
    y_c(nT) &= \int_{\infty}^{\infty}h_c(nT-\tau)u_c(\tau) d\tau \nonumber \\
    & = \sum_{k=-\infty}^{\infty}\int_{\infty}^{\infty}h_c(nT-\tau)\delta(\tau-kT) d\tau \, u(k) \nonumber \\
    & = \sum_{k=-\infty}^{\infty}h_c\left((n-k)T\right) u(k), 
    \label{eq:chain}
\end{align}
where the overall impulse response is given by the convolutions $h_c(t)=h_r(t)*h_o(t)*h_a(t)$. From \eqref{eq:invariance} and \eqref{eq:chain}, the output of a  discrete-time model with impulse response $h(n)$ is then
\begin{equation}
    y(n)=y_c(nT),
    \label{eq:model}
\end{equation}
as desired, for instance, in acoustic echo cancellation \cite{Azpicueta-Ruiz2011}. We note that \eqref{eq:model} is achieved even if there is aliasing due to $h_r(t)$ and $h_a(t)$ not being ideal low-pass filters.

\subsection{Realization}
Apart from aspects of numerical precision and computational complexity, the realization of $h(n)$ satisfying \eqref{eq:invariance} is simple. If the overall system with impulse response $h_c(t)$ is described by the state-space equations
\def\bxc{\mathbf{x}_c(t)}
\def\bxcd{\mathbf{x}'_c(t)}
\def\bx{\mathbf{x}}
\def\bA{\mathbf{A}}
\def\bF{\mathbf{F}}
\def\bbi{\mathbf{b}}
\def\bc{\mathbf{c}}
\begin{align*}
    \bxcd&=\bA\bxc+\bbi u_c(t) \\
    y_c(t)&=\bc^{\top} \bxc,
\end{align*}
then the discrete-time system realized by
\begin{align*}
\bx(n+1)&=e^{\bA T} \bx(n)+ e^{\bA T} \bbi u(n) \\
y(n)&=\bc^\top \bx (n) + \bc^\top \bbi u(n)
\end{align*}
has impulse response $h(n)$ satisfying \eqref{eq:invariance}.

\section{Generalization of impulse invariance to Volterra kernels} \label{sec:gen}

We assume now that the input/output relation of the analog portion of the chain in Fig. \ref{fig:chain} is given by the Volterra series  $y_c(t)= \sum_{p=1}^{\infty}y_{c,p}(t)$, with homogeneous outputs  given by
\begin{equation}\label{eq:ycp}
y_{c,p}(t)=  \int_{\mathbb{R}_+} \dots \int_{\mathbb{R}_+} h_{p}^{\text{tri}}(\tau_1,\ldots,\tau_p)\prod_{i=1}^{p}u_c(t-\tau_{i})d\tau_1\dots d\tau_p,
\end{equation}
where $h_{p}^{\text{tri}}(\tau_1,\ldots,\tau_p)$ is the triangular Volterra kernel\footnote{For simplicity, we omit the subscript $c$ for the continuous-time kernels.} of order $p$, which is null outside the domain $\tau_1 \le \ldots \le \tau_p$. The nonlinearity can arise, for instance, from a loudspeaker in acoustic echo cancellation.
Likewise, we assume now  that $y(n)=\sum_{p=1}^P y_p(n)$, where
\begin{equation}\label{eq:ypd}
    y_p(n)= \sum_{n_1=0}^{\infty}\ldots \sum_{n_p=0}^{\infty} v_p^{\text{tri}}(n_1,\ldots,n_p)\prod_{i=1}^p u(n-n_i),
\end{equation}
with $v_{p}^{\text{tri}}(n_1,\ldots,n_p)=0$ outside the domain $n_1 \le \ldots \le n_p$.

\subsection{Generalization of impulse invariance} 
Let $y_{c,p}(t)$ and $y_p(n)$ be given by \eqref{eq:ycp} and \eqref{eq:ypd}, respectively, and, as previously, $u_c(t)=\sum_n u(n)\delta(t-nT)$. As follows from \cite{Burt2018},
in order that $y_p(n)=y_{c,p}(nT)$ the impulse invariance relation \eqref{eq:invariance} has to be generalized to
\begin{equation} \label{eq:gen}
    v_p^{\text{tri}}(n_1,\ldots,n_p)=\frac{h_p^{\text{tri}}(n_1T,\ldots,n_pT)}{m_1 ! \ldots m_q !},
\end{equation}
$n_1\le\ldots\le n_p$, where $q$ is the number of distinct values among $n_1,\ldots,n_p$ and $m_1,\ldots,m_q$ are  their corresponding number of occurrences.\footnote{This result is stated without proof in \cite{Rugh1981}.} 
This modified relation is due to the possible discontinuity of the triangular continuous-time kernel on the border of the triangular domain $\tau_1 \le \dots \le \tau_p$.
In particular, in the interior of that domain we retrieve a direct extension of the invariance condition \eqref{eq:invariance}, that is, we have  $v_p^{\text{tri}}(n_1,\ldots,n_p)=h_p^{\text{tri}}(n_1T,\ldots,n_pT)$ for $n_1<\ldots < n_p$.

\subsection{Formulation for regular kernels}

For the analysis ahead, it will be more convenient to express $y_{c,p}(t)$ in terms of the regular kernels \cite{Rugh1981} \begin{equation} \label{eq:reg}
    h_p(\theta_1,\ldots,\theta_p)=h_p^{\text{tri}}(\tau_1,\ldots,\tau_p), 
\end{equation}
where 
\begin{equation} \label{eq:transf}
\theta_1=\tau_p-\tau_{p-1},\,\, \ldots, \,\,\theta_{p-1}=\tau_2 -\tau_1,\,\, \theta_p = \tau_1.    
\end{equation} 
With them we can write 
\begin{equation} \label{eq:hom}
   y_{c,p}(t) = \int_{\mathbb{R}_+} \!\!\dots \int_{\mathbb{R}_+} \!\!\!h_{p}(\theta_1,\ldots,\theta_p) \prod_{i=1}^p u(t-\bar{\theta}_i) d\theta_1\dots d\theta_p,
\end{equation}
where $\bar{\theta}_i = \sum_{j=1}^p \theta_j$. Likewise, in the discrete-time case,
\begin{equation}\label{eq:homdisc}
    y_p(n)= \sum_{n_p=0}^{\infty}\ldots \sum_{n_1=0}^{\infty} v_p(n_1,\ldots,n_p)\prod_{i=1}^p u(n-\bar{n}_i),
\end{equation}
where $\bar{n}_i=\sum_{j=1}^p n_j$.

We obtain now the condition corresponding to \eqref{eq:gen} for the regular kernels. Consider initially $p=4$ and some particular cases: \setlength\IEEEiedtopsep{1 mm}
\begin{itemize}

 \item $n_1,n_2,n_3>0$:  with $\theta_i=n_iT$ and \eqref{eq:transf} follows $\tau_1<\ldots<\tau_4$, so that from \eqref{eq:gen} and \eqref{eq:reg} follows  $v_p(n_1,\ldots,n_4)=h_p(n_1,\ldots,n_4)$ \vspace*{1mm}
 \item $n_1=0$; $n_2,n_3>0 \Rightarrow$  $\tau_1<\tau_2<\tau_3=\tau_4 \Rightarrow$ $v_p(0,n_2,n_3,n_4)=\frac{1}{2!}h_p(0,n_2,n_3,n_4)$\vspace*{1mm}
 \item $n_1,n_2=0$; $n_3>0 \Rightarrow$  $\tau_1<\tau_2=\tau_3=\tau_4 \Rightarrow$ $v_p(0,0,n_3,n_4)=\frac{1}{3!}h_p(0,0,n_3,n_4)$ \vspace*{1mm}
 \item $n_1,n_3=0$; $n_2>0 \Rightarrow$  $\tau_1=\tau_2<\tau_3=\tau_4 \Rightarrow$ $v_p(0,n_2,0,n_4)=\frac{1}{2!2!}h_p(0,n_2,0,n_4)$
 \end{itemize}
 The sought impulse invariance condition for regular kernels can then be seen to be
\begin{equation} \label{eq:genreg}
    v_p(n_1,\ldots,n_p)=\frac{h_p(n_1T,\ldots,n_pT)}{m_1 ! \ldots m_q !},
\end{equation}
where $q$ is  the number of  groups of consecutive null indices among $n_1,\ldots,n_{p-1}$ and $m_1-1,\ldots,m_q-1$ are the numbers of indices in each group.


\section{Realization}\label{sec:realiz}

\def\bH{\mathbf{H}} 
We consider here regular kernels given by the sum of separable factors  
\begin{equation}\label{eq:sep}
h_p(\tau_1,\ldots,\tau_p)=\sum_{r=1}^{R_p}\bH_r^{(p)}(\tau_p) \bH_{r}^{(p-1)}(\tau_{p-1})\ldots\bH_r^{(1)}(\tau_1),
\end{equation}for any set of matrix functions $\bH_r^{(i)}(\tau_i)$ of compatible dimensions. For simplicity, we refer to such kernels as separable kernels.
A particular case of \eqref{eq:sep} of great interest are the kernels of a bilinear system \def\bG{\mathbf{G}}
\begin{align*}
    \bxcd&=\bF\bxc+\bG\bxc u_c(t)+\bbi u_c(t) \\
    y_c(t)&=\bc^{\top} \bxc,
\end{align*}
which are given by \cite{Rugh1981}
\begin{equation*}
 { h_p(\tau_1,\ldots,\tau_p)}
 = \bc^{\T}e^{\bF \tau_p} \bG e^{\bF \tau_{p-1}} \bG \ldots \bG e^{\bF \tau_1} \bbi ,\,\, \tau_i \ge 0. \label{eq:kbilin}
\end{equation*}
For $p>2$ we can assign then $\bH^{(1)}(\tau_1)=e^{\bF \tau_1}\bbi$, $\bH^{(i)}(\tau_i)=e^{\bF \tau_i}\bG$,  $1<i<p$, and  $\bH^{(p)}(\tau_p)=\bc^\top e^{\bF \tau_p}\bG$. In this case, $R_p=1$ so the subscripts in $\bH_r^{(i)}(\tau_i)$ were omitted.
Without loss of generality, we consider $R_p=1$ from here onward.

\subsection{Cascade structure}
\begin{figure}
    \centering
    \includegraphics{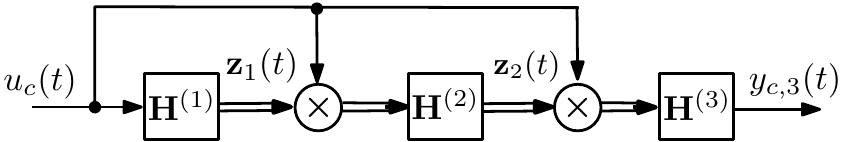}
    \caption{Cascade realization of a separable kernel, $p=3$.}
    \label{fig:cascade}
\end{figure}

 From \eqref{eq:sep} and \eqref{eq:hom} it follows that separable kernels can be realized quite simply by a cascade of linear blocks and multipliers, as depicted in Fig. \ref{fig:cascade} for $p=3$, where
\def\bz{\mathbf{z}}
\begin{align*}
\bz_1(t)&=\int_0^\infty \bH^{(1)}(\tau_1)u_c(t-\tau_1)d\tau_1, \\
\bz_2(t)&=\int_0^\infty \bH^{(2)}(\tau_2)\bz_1(t-\tau_2)u_c(t-\tau_2)d\tau_2, \\ y_{c,p}(t)&=\int_0^\infty \bH^{(3)}(\tau_3)\bz_2(t-\tau_3)u_c(t-\tau_3)d\tau_3.
\end{align*}
It should be noted that this also would be a realization of any kernel equivalent to the regular separable kernel $h_p(\tau_1,\ldots,\tau_p)$, such as for instance the triangular kernel $h_p^{\text{tri}}(\tau_1,\ldots,\tau_p)=h_p(\tau_p-\tau_{p-1},\ldots, \tau_2-\tau_1,\tau_1)$, which in general is not separable.

In the discrete-time case, we note initially that, for the cascade structure in particular, replacing $\bH^{(i)}(\tau_i)$ with $\bH^{(i)}(n_iT)$ is a realization \footnote{This is not true in general. For instance, the series connection of discrete-time linear systems $f(n)=f_c(nT)$ and $g(n)=g_c(nT)$ has impulse response $\sum_k f(k)g(n-k) \neq \int f_c(\tau)g_c(nT-\tau)d\tau$.} of $v_p(n_1,\ldots,n_p)=h_p(n_1T,\ldots,n_pT)$.  This, in general, does not satisfy \eqref{eq:genreg}, so the goal $y_p(n)=y_{c,p}(nT)$ for $u_c(t)=\sum_n u(n)\delta(t-nT)$ would not be achieved.
Another way of viewing the problem follows from rewriting  \eqref{eq:genreg} as \begin{equation}\label{eq:mfunc}
    v_p(n_1,\ldots,n_p)=m(n_1,...,n_{p-1})h_p(n_1T,\ldots,n_pT),
\end{equation}
where $m(n_1,...,n_{p-1})$ implements the described rule for the factor $1/m_1 ! \ldots m_q!$. This emphasizes the fact that the discrete-time kernels are not separable and, therefore, their realization is not simply a discrete-time version of the cascade structure type depicted in Fig. \ref{fig:cascade}. Nevertheless, as shown in the following their realization is less complicated than may seem initially.

\subsection{Cascade operator}
In order that the following analysis be more compact, we define, given $u(n)$, the \emph{cascade operation} on a signal $x(n)$ as
\begin{equation*}
    v_p \circ x(n) = \sum_{n_p=0}^{\infty}\!\!\ldots \!\! \sum_{n_1=0}^{\infty} v_p(n_1,\ldots,n_p)x(n-\bar{n}_i)\prod_{i=2}^p u(n-\bar{n}_i),
\end{equation*}
which for $p=1$ is the convolution operation $v_p * x(n)=\sum_k v_p(k)x(n-k)$.
We can rewrite \eqref{eq:homdisc} then as 
  $y_p(n)=v_p\circ u(n)$.
The properties below follow directly from the definition and will be useful. The first two mean that the operation is bilinear.
\begin{itemize}
    \setlength\itemsep{1mm}
    \item $v_p(n_1,\ldots,n_p)=r(n_1,\ldots,n_p)+s(n_1,\ldots,n_p)\Rightarrow \\ v_p\circ x(n) = r\circ x(n)+s\circ x(n)$
    \item $v_p\circ [x(n)+y(n)]=v_p\circ x(n)+ v_p \circ y(n)$
    \item $ v_p(n_1,\ldots,n_p)=v^{(1)}(n_1)v^{(2)}(n_2,\ldots,n_p)\Rightarrow  v_p\circ x(n) = v^{(2)}\circ x_1(n)$, where $x_1(n)=[v^{(1)} * x(n)]u(n)$
\end{itemize}

\subsection{Corrections for impulse invariance}
Without loss of generality,  we consider, for notational simplicity, the case of kernels separable into scalar (instead of matrix) factors, $ h_p(\tau_1,\ldots,\tau_p)=h^{(1)}(\tau_1)\ldots h^{(p)}(\tau_p)
$. We also omit the sampling period $T$ in \eqref{eq:mfunc} and write
\begin{align} \label{eq:discret}
    v_p(n_1,\ldots,n_p)
    &=m(n_1,\ldots,n_{p-1})h^{(1)}(n_1)\ldots h^{(p)}(n_p).
\end{align}
Finally, we will make use of $m(n_1,\ldots,n_\ell)$, with $\ell<p-1$, for which the same definitions apply.

In the following, we progressively expand the analysis to cover the cases $p=2,3$ and 4, after which the realization of impulse invariant separable kernels for any $p$ will become clear.
 \subsubsection{Case $p=2$} 
\begin{figure}
    \centering
    \includegraphics{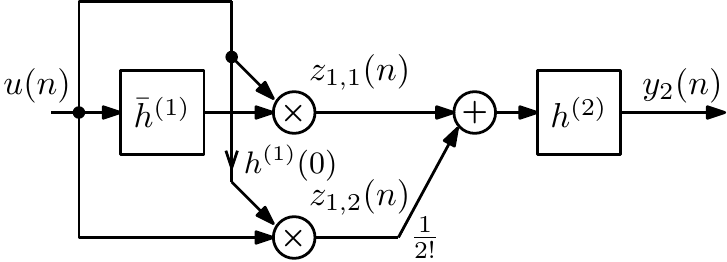}
    \caption{Realization of impulse invariant kernel of order  $p=2$.}
    \label{fig:realiz2}
\end{figure}

Placing a null sample at the origin of an impulse response, we define
\begin{equation*}
    \bar{h}^{(i)}(n_i)=[1-\delta(n_i)]h^{(i)}(n_i)
\end{equation*}
From   \eqref{eq:discret} we can always write then
$ v_p(n_1,n_2)=r(n_1,n_2)+s(n_1,n_2),
$ where
\begin{align}\label{eq:defr}
    r(n_1,n_2) &= \bar{h}^{(1)}(n_1)r^{(2)}(n_2) \\\label{eq:defs}
    s(n_1,n_2)&=h^{(1)}(0)\delta(n_1)s^{(2)}(n_2),
\end{align}
with 
$r^{(2)}(n_2)=h^{(2)}(n_2)$ and
$s^{(2)}(n_2)=m(0)h^{(2)}(n_2)$.
It follows then that
\begin{align}\label{eq:yp2}
    y_p(n)&=h^{(2)} * z_{1,1}(n)+h^{(2)} * \left[ m(0)z_{1,2}(n)\right]\\
    &=h^{(2)}* \left[z_{1,1}(n)+ m(0)z_{1,2}(n)\right],
\end{align}
 where $m(0)=1/2$ and
\begin{align}\label{eq:defz1}
    z_{1,1}(n)&=\left[\bar{h}^{(1)}* u(n)\right]u(n) \\
    \label{eq:defz2}
    z_{1,2}(n)&={h}^{(1)}(0) \, u(n) \, u(n)
\end{align}
This realization is depicted in Fig.~\ref{fig:realiz2}.

\subsubsection{Case $p=3$} From
$ v_p(n_1,n_2,n_3)=r(n_1,n_2,n_3)+s(n_1,n_2,n_3)$ now, \eqref{eq:defr} and \eqref{eq:defs} become \begin{align*}
    r(n_1,n_2,n_3) &= \bar{h}^{(1)}(n_1)r^{(2)}(n_2,n_3)\\
    s(n_1,n_2,n_3)&=h^{(1)}(0)\delta(n_1)s^{(2)}(n_2,n_3),
\end{align*}
where 
$r^{(2)}(n_2,n_3)=m(n_2)h^{(2)}(n_2)h^{(3)}(n_3)$ and $s^{(2)}(n_2,n_3)=m(0,n_2)h^{(2)}(n_2)h^{(3)}(n_3)$.
We see that $m(n_2)$ and $m(0,n_2)$ don't decouple from $h^{(2)}(n_2)$, so \eqref{eq:yp2} has to be written now as
\begin{equation}\label{eq:yp3}
    y_p(n)=r^{(2)}\circ z_{1,1}(n)+s^{(2)}\circ z_{1,2}(n),
\end{equation}
where, nevertheless, \eqref{eq:defz1} and \eqref{eq:defz2} still apply.

\begin{figure*}
    \centering
    \includegraphics{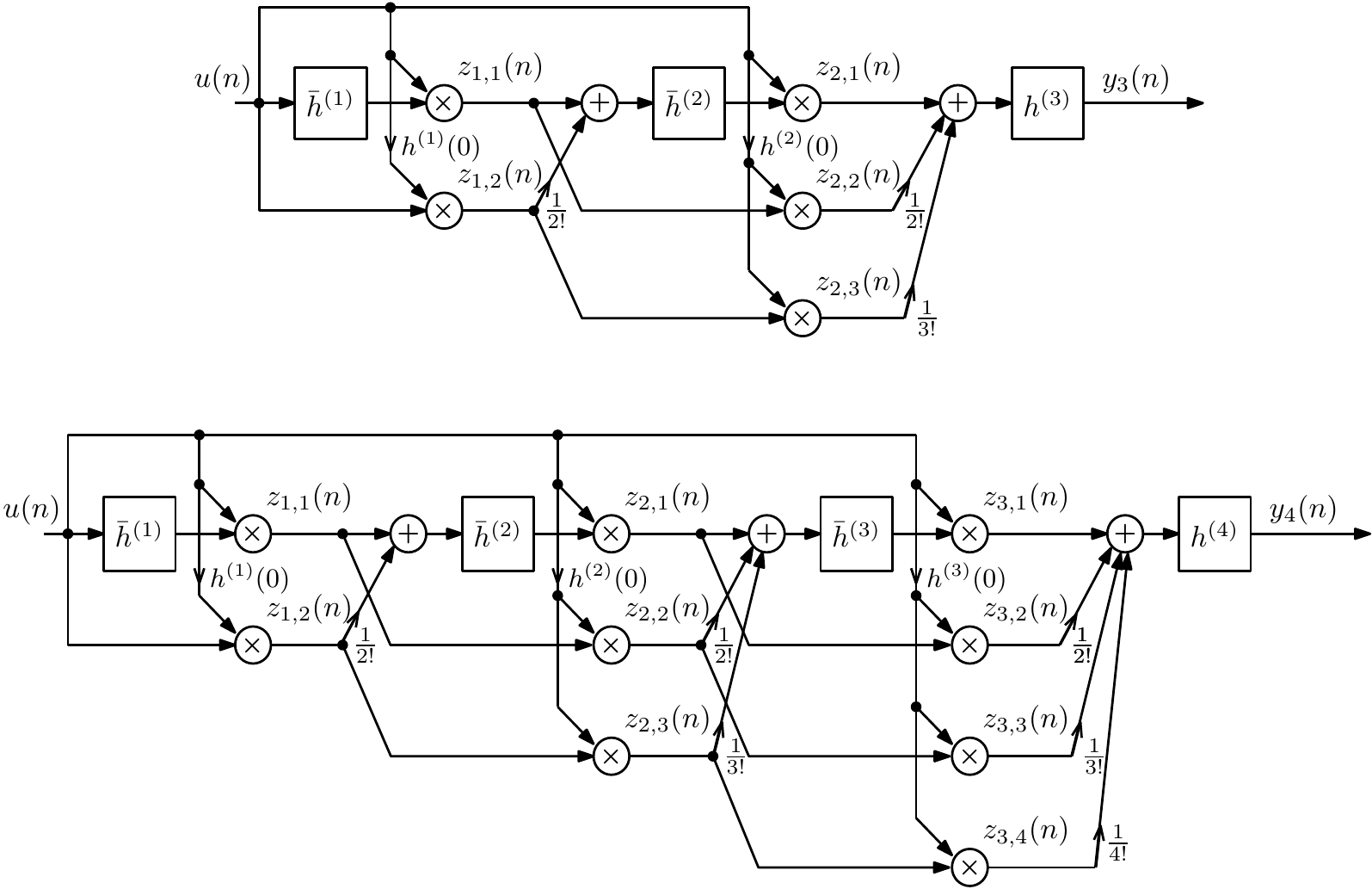}
    \caption{Realization of impulse invariant separable kernels of orders $p=3$ and $p=4$.}
    \label{fig:realiz3}
\end{figure*}

Noting that $m(n_2)=1$ and $m(0,n_2)=m(0)$ if $n_2>0$, we can then rewrite 
\begin{multline*}
r^{(2)}(n_2,n_3) =\bar{h}^{(2)}(n_2) h^{(3)}(n_3) \\ +m(0)h^{(2)}(0)\delta(n_2)h^{(3)}(n_3),
\end{multline*}
\begin{multline*}
s^{(2)}(n_2,n_3)=m(0)\bar{h}^{(2)}(n_2) h^{(3)}(n_3)\\ +m(0,0)h^{(2)}(0)\delta(n_2)h^{(3)}(n_3).
\end{multline*}
Inserting these expressions into \eqref{eq:yp3}, we get
\begin{multline}\label{eq:yp3b}
    y_p(n)=h^{(3)}* z_{2,1}(n)+ h^{(3)}* \left[ m(0)z_{2,2}(n)\right]+\\h^{(3)}* \left[m(0,0)z_{2,3}(n)\right] \\
    =h^{(3)}* \left[z_{2,1}(n)+ m(0)z_{2,2}(n)+m(0,0)z_{2,3}(n)\right],
\end{multline}
where $m(0,0)=1/3!$ and
\begin{align}\label{eq:defz21}
    z_{2,1}(n)&=\left[\bar{h}^{(2)}* [z_{1,1}(n)+m(0)z_{1,2}(n)]\right]u(n), \\
    \label{eq:defz22}
    z_{2,2}(n)&=h^{(2)}(0)z_{1,1}(n)u(n), \\
    \label{eq:defz23}
    z_{2,3}(n)&=h^{(2)}(0)z_{1,2}(n)u(n). \end{align}
This realization is depicted in Fig.~\ref{fig:realiz3}.

\subsubsection{Case $p=4$} We can start from \eqref{eq:yp3}, which still applies, but now with
\begin{align*}r^{(2)}(n_2,n_3,n_4) &=m(n_2,n_3) h^{(2)}(n_2)t(n_3,n_4) \\
s^{(2)}(n_2,n_3,n_4) &=m(0,n_2,n_3) h^{(2)}(n_2)t(n_3,n_4),
\end{align*}
where $ t^{(3)}(n_3,n_4)\triangleq h^{(3)}(n_3)h^{(4)}(n_4)$. Since  $m(n_2,n_3)$ and $m(0,n_2,n_3)$ do not decouple from $t^{(3)}(n_3,n_4)$, \eqref{eq:yp3b} becomes, with a slight abuse of notation in favor of expediency,
\begin{multline*}\label{eq:yp4}
    y_p(n)=\left[m(n_3)t^{(3)}\right]\circ z_{2,1}(n)+\left[m(0,n_3)t^{(3)}\right]\circ z_{2,2}(n)  \\ + \left[m(0,0,n_3)t^{(3)}\right]\circ z_{2,3}(n),
\end{multline*}
where \eqref{eq:defz21}, \eqref{eq:defz22} and \eqref{eq:defz23} still apply. Substituting now in the expression above
\begin{multline*}
    m(n_3)t^{(3)}({ n_3,n_4})=\bar{h}^{(3)}(n_3)h^{(4)}(n_4)\\ +m(0)h^{(3)}(0)\delta(n_3)h^{(4)}(n_4)
\end{multline*}
\begin{multline*}
    m(0,n_3)t^{(3)}(n_3,n_4)=m(0)\bar{h}^{(3)}(n_3)h^{(4)}(n_4)\\ +m(0,0)h^{(3)}(0)\delta(n_3)h^{(4)}(n_4)
\end{multline*}
\begin{multline*}
    m(0,0,n_3)t^{(3)}(n_2,n_3)=m(0,0)\bar{h}^{(3)}(n_3)h^{(4)}(n_4)\\ +m(0,0,0)h^{(3)}(0)\delta(n_3)h^{(4)}(n_4),
\end{multline*}
we get
\begin{multline*}
    y_p(n)=h^{(4)}* \left[z_{3,1}(n)+ m(0)z_{3,2}(n)+m(0,0)z_{3,3}(n) \right.\\ \left. +m(0,0,0)z_{3,4}(n)\right],
\end{multline*}
where $m(0,0,0)=1/4!$,
\begin{multline*}
    z_{3,1}(n)=[\bar{h}^{(3)}* [z_{2,1}(n)+m(0)z_{2,2}(n)\\   +m(0,0)z_{2,3}]]u(n)
\end{multline*}\begin{align*} z_{3,2}(n)&=h^{(3)}(0)z_{2,1}(n)u(n) \\  z_{3,3}(n)&=h^{(3)}(0)z_{2,2}(n)u(n)\\ z_{3,4}(n)&=h^{(3)}(0)z_{2,3}(n)u(n) \end{align*}
This realization is also depicted in Fig. \ref{fig:realiz3}.

\subsubsection{Generalization}
From the above, we can see that the proposed  realization of an impulse invariant  Volterra kernel of order $p>1$, \[v_p(n_1,\ldots,n_p)=h_p(n_1T_c,\ldots,n_pT_c)/(m_1!,\ldots,m_q!),\]where $h_p(\tau_1,\dots,\tau_p)$ is separable, consists of the following steps: 
\begin{itemize}
    \item For $i=1,\ldots,p-1$ and with $z_{0,1}(n)\triangleq u(n)$, compute 
 \begin{align*}z_{i,1}(n)&=\left[\bar{h}^{(i)}*\sum_{j=1}^{i}\frac{1}{j!}z_{i-1,j}(n)\right] u(n),\\
    z_{i,j}(n)&=h^{(i)}(0)z_{i-1,j-1}(n)u(n), \,\, j=2,\ldots,i+1.\end{align*}
    \item Compute
    $\displaystyle y_p(n)=h^{(p)}*\sum_{j=1}^{p}\frac{1}{j!}z_{p-1,j}(n).$
\end{itemize}

\section{Computational complexity} \label{sec:complex}

In the following, for a separable kernel $h_p(\tau_1,\ldots,\tau_p) = \bH^{(p)}(\tau_p)\ldots \bH^{(1)}(\tau_1)$, we calculate the additional multiplications required to realize the impulse invariant discrete-time kernel $v_p(n_1,\ldots,n_p)=h_p(n_1T,\ldots,n_pT)/(m_1!,\ldots,m_q!)$ as opposed to simply realizing $v_p(n_1,\ldots,n_p)=h_p(n_1T,\ldots,n_pT)$ with the structure in Fig. \ref{fig:cascade}.

\subsection{Scalar case}
We consider initially that $\bH^{(i)}(\tau_i)=h^{(i)}(\tau_i)$ are scalar impulse responses. We assume also that the product $h^{(i)}(0)u(n)\triangleq w_i(n)$ in each stage does not add to complexity, since it is compensated by the direct, series or paralell realization of the rational system $\bar{h}^{(i)}$ requiring one multiplication less than that of $h^{(i)}$. So at the $i$th stage, $i < p-1$, there are the $2i$ additional multiplications required by the operations $w_i(n)z_{i-1,j-1}(n)(1/j!)$, $j=2,\ldots,i+1$, totaling $2\sum_{i=1}^{p-2} i = (p-1)(p-2)$ multiplications. At the $(p-1)$th stage, one can first perform $\sum_{j=2}^{p} z_{p-1,j}(n)(1/j!)$ and then multiply this by $w_i(n)$, so only $p$ additional multiplications are performed, instead of $2(p-1)$. The overall number $A_S$ of additional multiplications is, therefore, $(p-1)(p-2)+p=p(p-2)+2$:
\begin{equation}
     A_S=p(p-2)+2
\end{equation}

\subsection{General case}
We consider now the more general case $h_p(\tau_1,\ldots,\tau_p) = \bH^{(p)}(\tau_p)\ldots \bH^{(1)}(\tau_1)$, with  matrices $\bH^{(i)}(\tau_i)$ having dimensions $M_i\times M_{i-1}$, where $M_p=M_0=1$. \def\bw{\mathbf{W}}\def\bz{\mathbf{z}}  Here, a state space-space realization of a rational system $\bar{\bH}^{(i)}$ may have, in some cases, the same complexity of that of $\bH^{(i)}$. So we take into account  the product  $\bH^{(i)}(0)u(n)\triangleq \bw_i(n)$, which requires $M_i M_{i-1}$ multiplications.
In turn, for $i<p-1$, the products $\bw_i(n)\bz_{i-1,j-1}(n)=\bz_{i,j}(n)$ and $\bz_{i,j}(n)(1/j!)$, $j=2,\ldots,i+1$ require, respectively, $M_i M_{i-1}$ and $M_i$ multiplications for each value of $j$. If $\bH^{(i)}(0)$ has a known structure (such as when the bilinear system is obtained from Carleman bilinearization \cite{Rugh1981} of a linear-analytic system) then the first product (and also $ \bH^{(i)}(0)u(n))$ require $\mu_i$ multiplications, where $\mu_i\le M_i M_{i-1}$ is the number of non-null elements of  $\bar{\bH}^{(i)}$. Up to stage $p-2$ then, a total of $ \sum_{i=1}^{p-2} \mu_i+i(\mu_i + M_i)$ multiplications is required. Similarly to the scalar case, at the $(p-1)$th stage, $ p>2$,  one can first perform $\sum_{j=2}^{p} \bz_{p-1,j}(n)(1/j!)$, then multiply this by $\bH^{( p-1})(0)$ and finally by $u(n)$, so only $\mu_{p-1}+ pM_{p-1}$ multiplications are performed, instead of $\mu_{p-1}+(p-1)(\mu_{p-1}+M_{p-1})$. The overall number $A_M$ of additional multiplications is therefore $ \mu_{p-1}+pM_{p-1} + \sum_{i=1}^{p-2}\mu_i+i(\mu_i +M_i)$, $ p>2$.For $p=2$, $\bH^{(1)}(0)$ can absorb the factor $1/2$, so the overall number of additional multiplications is $\mu_1+M_1$. So, in synthesis,
\begin{equation}
    A_M = \left\{ \begin{array}{cc}
        \mu_1+M_1, & p=2 \\
         \mu_{p-1}+pM_{p-1}+\sum_{i=1}^{p-2}\mu_i+i(\mu_i +M_i), & p>2
    \end{array}\right.
\end{equation}
\section{Conclusion}
We have shown how the generalized impulse invariance condition for triangular Volterra kernels translates to regular kernels. Such kernels, if separable, have simple continuous-time realizations with a cascade structure. By defining a cascade operator, we have shown how such structures should be modified for the discrete-time realization of impulse invariant  separable kernels. Finally, we assessed the additional computational complexity incurred by that modification.

\bibliographystyle{./bibliography/IEEEtran}
\bibliography{./bibliography/goulart-burt,./bibliography/IEEEabrv}

\end{document}